%% file: main.tex
\documentclass[10pt,journal,compsoc]{IEEEtran}

\usepackage[utf8]{inputenc}
\usepackage[T1]{fontenc}
\usepackage{amsmath,amssymb,amsfonts}
\usepackage{graphicx}
\usepackage{booktabs}
\usepackage{array}
\usepackage{makecell}
\usepackage{tabularx}
\usepackage{xcolor}
\usepackage{listings}
\usepackage{cite}
\usepackage[hyphens]{url}
\usepackage{hyperref}

\definecolor{googleblue}{HTML}{1A73E8}
\definecolor{neutralgray}{HTML}{5F6368}
\definecolor{codebg}{HTML}{F4F6F8}
\hypersetup{colorlinks=true, linkcolor=googleblue, citecolor=googleblue, urlcolor=googleblue,
  pdftitle={Connection Escrow: Holding AI-Agent Pod Egress in the Kernel Until an Isolated Verifier Rules},
  pdfauthor={Ashok Subbabhatta Gopalakrishna}}
\begin{document}

\title{Connection Escrow: Holding AI-Agent Pod Egress in the Kernel Until an Isolated Verifier Rules}

\author{Ashok~Subbabhatta~Gopalakrishna%
\IEEEcompsocitemizethanks{\IEEEcompsocthanksitem Independent researcher. E-mail: sg.ashok@gmail.com. Code: \url{https://github.com/45h0kg/benign-attach-gateway}.}}

\IEEEtitleabstractindextext{%
\begin{abstract}
AI agents running in containers act on the world by opening network connections, and the text that drives them can be hostile. Controls placed between the agent and its tools, such as gateways, SDK hooks and proxy settings, can be bypassed by a child process or a direct socket, and even when they are not bypassed, their approval is not tied to the connection the kernel later opens. We present \emph{connection escrow}: every outbound TCP connection from a supervised pod is parked inside the kernel's \texttt{connect()} path, before any packet exists, by a 98-instruction eBPF program attached to the pod's cgroup. An interceptor on the node attributes the held socket to its kernel socket cookie, process, executable digest, cgroup and pod identity, parks the first bytes the agent writes (HTTP request line or TLS ClientHello), and asks a verifier that runs in a separate container with its own key. The connection is released only on a signed verdict that names that one socket cookie and destination, is unexpired and unused; no verdict by the deadline resets the connection, and a late approval is refused. The agent is unmodified. We implement the system (Python plus eBPF, with an iptables fallback) and evaluate it two ways. A 36-check adversarial suite on Linux 6.18 with real cgroups and uid-separated agent processes passes on both hold points (the three UDP-pinning checks need the kernel path): refused connections never reach the destination, forged, rebound, replayed, expired and late verdicts are refused, the agent's own attempt to reach the verifier is escrowed and refused, child-process and renamed-binary bypasses fail, and UDP is pinned to the cluster resolver. On a single-node Kubernetes cluster, an unmodified agent pod is held per connection and attributed to its SPIFFE identity; a raw socket that bypasses a comparison sidecar proxy is held by escrow; and six pods opening 600 connections are all gated. The hold adds 4.3\,ms median (6.8\,ms p95) per released connection on loopback, of which 24\,$\mu$s is policy evaluation; the rest is attribution, one HTTP round trip and two signatures. Finally, we extend escrow to gate on request \emph{content} without terminating TLS: an eBPF uprobe reads every plaintext write at the agent's TLS-library boundary, a model scores it, and each TLS record is held until the write that produced it is scored, while the TLS session stays end-to-end to the real server with the real certificate and the enforcer holds no key. The gate is demonstrated on three runtime shapes with two kernel programs --- shared OpenSSL, statically linked OpenSSL (the BoringSSL shape) and Go's \texttt{crypto/tls} --- against the keep-alive exploit that defeats a first-record check: on one connection a benign request is allowed and an exfiltration request that follows it is denied on its captured plaintext, so the server receives only the first. What decides a verdict is deliberately pluggable and out of scope.
\end{abstract}
}

\maketitle
\IEEEdisplaynontitleabstractindextext
\IEEEpeerreviewmaketitle

\section{Introduction}
\IEEEPARstart{A}{n} AI agent that can call tools can open network connections: to fetch a page, to post to an API, to close a ticket, to push a commit. The decision to do so comes from text the agent was given, and some of that text is written by whoever wants the agent to misbehave. The question this paper asks is narrow: once the agent has decided to connect, where can the decision be checked by something the agent cannot influence or bypass, and how can the check be bound to the connection that actually happens?

Today's answers sit at the application layer. A gateway inspects tool calls; an SDK hook vetoes them; a sandbox sets a proxy environment variable. Each is bypassed by a tool that opens its own socket, by a child process, or by a runtime that ignores the proxy setting. Kernel-level policy closes that gap but, in the systems we know of, applies a standing allow-list: once a destination is permitted, it is permitted for every connection, by every process in scope, for as long as the policy is loaded. Nothing ties a particular approval to a particular socket.

We propose to \emph{escrow} the connection. The kernel parks it at \texttt{connect()}, before a SYN exists; a node-local interceptor attributes it to the socket, the process and the pod; a verifier in a separate container rules; and the interceptor releases exactly the socket the verdict names, once. If no verdict arrives in time, the connection is reset. The verifier is a destination like any other, so the agent cannot reach it: its attempt is escrowed and refused.

The contributions are:
\begin{enumerate}
  \item \textbf{A kernel hold point for pod egress.} A \texttt{cgroup/connect4} eBPF program records the socket cookie, original destination, cgroup id and pid of every TCP \texttt{connect()} from the supervised cgroup and rewrites the destination to the interceptor (Section~\ref{sec:hold}). An iptables \texttt{REDIRECT} keyed on the agent's uid gives the same hold point without eBPF.
  \item \textbf{An attribution record that joins the kernel view to the platform view.} Socket cookie, inode and uid from \texttt{inet\_diag}; pid, cgroup, executable digest and \texttt{comm} from the eBPF map and \texttt{/proc}; pod uid, namespace, service account and SPIFFE id from a registry keyed on the cgroup path (Section~\ref{sec:attrib}).
  \item \textbf{A release gate bound to one socket.} A verdict is accepted only if it is signed by the verifier, names the escrow id, the socket cookie and a hash of the destination and the parked first bytes, is unexpired and unused, and arrives before the deadline (Section~\ref{sec:gate}). Hold-and-release lets an operator decide later, inside the deadline (Section~\ref{sec:hold-release}).
  \item \textbf{Content gating of encrypted connections without terminating TLS (Section~\ref{sec:content}).} An eBPF uprobe on the agent's TLS library --- shared OpenSSL, statically linked OpenSSL/BoringSSL, or Go's \texttt{crypto/tls}, two kernel programs in all --- reads every plaintext write at the library boundary, before encryption; the interceptor holds each application-data record until the write that produced it is scored, so a second request on a kept-alive connection is gated like the first; and the connection stays end-to-end to the real server, with the real certificate, the enforcer holding no key. This gates on \emph{what} the agent sends, not only \emph{where}, while keeping the tunnel intact.
  \item \textbf{An implementation and a measured evaluation.} Python and eBPF, 36 adversarial checks on the eBPF hold point (33 of which apply to iptables), a latency breakdown with and without native cryptography, a Kubernetes run, and the content-gate demonstration (Section~\ref{sec:eval}). Everything reported comes from logged runs of \texttt{experiments/run\_all.sh} and the cluster drivers.
\end{enumerate}

What is \emph{not} in this paper: the logic that decides whether a connection is acceptable. The verifier ships with a rule engine (destination by SNI, host or address; port; HTTP method and path prefix; per-workload budgets; operator hold), and the content gate ships with a placeholder classifier. Any function from an attribution record, or a captured request, to a verdict can replace either. The paper is about where the check runs and what binds its answer to the connection.

\section{Problem and threat model}
\label{sec:problem}

\subsection{Setting}
A \emph{supervised workload} is a Kubernetes pod (or, more generally, any control group) in which an agent and the tool processes it starts run. The pod has a service account and, through it, a platform identity such as a SPIFFE verifiable identity document (SVID). The operator wishes to control every outbound connection the pod makes, one connection at a time, according to a policy that a \emph{verifier} evaluates. The verifier is trusted by the operator and is outside the agent's control; it runs elsewhere in the cluster, in its own pod with its own key. The question is not what the policy should say --- that is the operator's and is deliberately out of scope --- but \emph{where} the decision can be enforced so that the agent cannot route around it, and \emph{how} the decision can be tied to the exact connection it is about.

This matters because an agent's outbound calls are the point at which text it was given turns into action in the world: fetching a URL named in a web page, posting to an API named in a tool result, pushing to a host named in a prompt. When that text is adversarial --- a prompt injection, a poisoned document, a compromised upstream tool --- the connection the agent opens is the thing that must be checked, and checked somewhere the agent's own code cannot disable.

\subsection{Assets and trust boundary}
The assets to protect are (i) the confidentiality of data the pod holds, which an unapproved outbound connection could exfiltrate, and (ii) the integrity of external systems, which an unapproved connection could act on (write, delete, transfer). The \emph{trusted computing base} is the operating-system kernel, the node agent that loads the eBPF programs and runs the interceptor, the verifier, and the cluster control plane (API server, kubelet, CNI). The \emph{untrusted} part is everything inside the supervised pod: the agent process, every process it spawns, every library and interpreter it loads, and any tool server it starts. The boundary between the two is the pod's cgroup: the kernel is on one side, the untrusted workload on the other, and the eBPF hook sits exactly on that line, inside the kernel, invoked by the untrusted code but not modifiable by it.

\subsection{Adversary capabilities}
We grant the adversary full control of the code running inside the pod. Concretely, the adversary may: open a connection to any address, port and protocol; issue the connection from the agent process, from a child process, from a thread, or from a freshly \texttt{exec}'d binary; rename, copy or rebuild its executables to defeat path-based rules; craft arbitrary application-layer bytes (any HTTP request, any TLS ClientHello, any opaque stream); attempt to contact the verifier directly to forge or solicit an approval; replay a verdict it has observed; and open many connections concurrently to race the control. The adversary may observe traffic on the node network between the interceptor and the verifier. The adversary may \emph{not} escape the pod's cgroup (that is container escape, a separate problem assumed solved by the sandbox), obtain root on the node, read the verifier's signing key or the interceptor's record-authentication key, or forge the kernel's socket-cookie allocation. We assume the sandbox and the kernel's namespace isolation hold; escrow composes with them rather than replacing them.

\subsection{Goals}
\label{sec:goals}
Against that adversary the system must meet six goals. Each is stated as a property and the reason it is hard without a kernel-resident, per-connection mechanism.

\noindent\textbf{G1 --- Pre-packet hold.} A connection the policy refuses must produce \emph{no packet} to the destination, not merely a reset after the SYN. A control that acts after the connection is established (a proxy that reads the request, a filter that sees the first packet) has already leaked the fact of the attempt, and sometimes the first bytes, to the destination. Holding inside \texttt{connect(2)}, before a route is allocated, is what makes the refusal invisible to the outside.

\noindent\textbf{G2 --- Attribution.} Every held connection must be attributed to the kernel object that made it (the socket, identified by its cookie), to the process (pid, executable digest, cgroup), and to the platform identity (pod, namespace, service account, SPIFFE id). Without attribution the verifier cannot apply a per-workload policy, and a receipt cannot be trusted to name who did what. Attribution must survive the adversary renaming its binary or acting from a child process.

\noindent\textbf{G3 --- Isolated verifier.} The verdict must come from a component the agent can neither reach nor impersonate. If the agent could open a connection to the verifier, it could request its own approvals; if it could forge a record, it could answer as the interceptor. Both paths must be closed.

\noindent\textbf{G4 --- Single-use, socket-bound release.} An approval must release exactly one socket, to exactly one destination, exactly once. This is what closes the time-of-check-to-time-of-use (TOCTOU) gap: an approval that authorises ``connections to host $H$'' can be spent by a later, different connection to $H$; an approval bound to one socket cookie and one destination hash, consumed atomically, cannot. Replay of a captured verdict, rebinding it to another socket or destination, and forgery under the wrong key must all fail.

\noindent\textbf{G5 --- Fail closed.} If no verdict arrives before a deadline, or the verifier is unreachable, the connection must be reset, not allowed. An approval that arrives \emph{after} the deadline must have no effect, so that a slow or replayed verdict cannot open a connection the agent attempted earlier and the system cannot be pushed into a fail-open state by delaying the verifier.

\noindent\textbf{G6 --- No agent change.} The control must require nothing of the agent: no SDK, no proxy environment variable, no modified library, no injected sidecar in the agent's own process. Anything the agent must opt into, it can opt out of; and anything that lives in the agent's address space, a compromised agent can disable. The enforcement must hold for an unmodified binary, in any language, including code the agent writes and runs at run time.

\subsection{Non-goals}
Escrow does not decide \emph{whether} a connection is acceptable; that is the policy's job, and the verifier's policy engine is replaceable (Section~\ref{sec:verifier}). It does not defend against an adversary who has escaped the pod or holds root on the node. Without the content gate it does not inspect or control traffic on connections it has released --- once a socket is spliced to its destination, the bytes flow unexamined beyond the first bytes checked at release; with the content gate every write on the connection is scored (Section~\ref{sec:content}). It does not provide confidentiality for the traffic itself (that is TLS's job) and does not terminate TLS. Finally, it is not a replacement for network policy or a service mesh; it composes with them, adding a per-connection, identity-bound, fail-closed decision that those layers do not provide.

\section{Design}
\label{sec:design}
\begin{figure*}[t]
\centering
\includegraphics[width=0.95\textwidth]{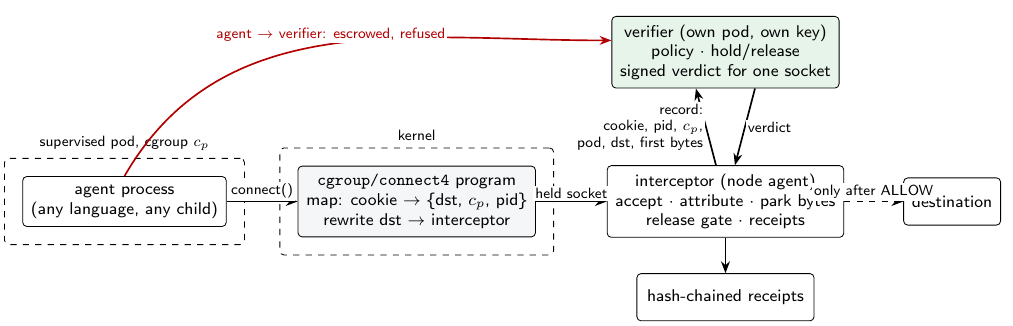}
\caption{Connection escrow. The agent's \texttt{connect()} is rewritten in the kernel to the interceptor and recorded under its socket cookie. The interceptor attributes the held socket, parks the first bytes, and asks the verifier; only an ALLOW bound to that socket lets it connect to the destination. An agent's attempt to reach the verifier is itself escrowed and refused.}
\label{fig:arch}
\end{figure*}

\begin{figure*}[t]
\centering
\includegraphics[width=0.96\textwidth]{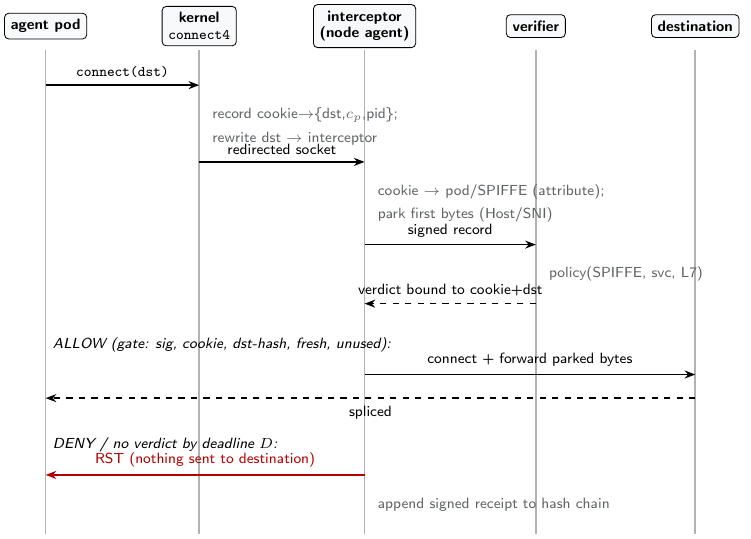}
\caption{One escrowed connection, across the five components. The kernel records the socket cookie and rewrites the destination; the interceptor attributes the redirected socket and parks the first bytes; the verifier returns a verdict bound to that cookie and destination. On ALLOW the interceptor connects out and splices; on DENY or deadline it resets, and nothing reaches the destination.}
\label{fig:seq}
\end{figure*}

Fig.~\ref{fig:arch} shows the components, Fig.~\ref{fig:seq} the exchange between them for one connection, and Fig.~\ref{fig:timeline} the life of a connection against the deadline. The design follows from a single commitment: the decision point is the kernel's \texttt{connect()} path, and everything else --- attribution, the verifier call, the release --- hangs off the socket that path produces.

\subsection{Why the \texttt{connect()} hook}
\label{sec:rationale}
Several places could host an egress control, and the choice determines which goals are reachable. A \emph{sidecar or egress proxy} (Envoy \texttt{ext\_authz}, a CONNECT proxy) sees only traffic the workload chooses to send it, so it fails G6 and G1: a child process or a raw socket bypasses it, and by the time it sees a request a connection already exists. A \emph{netfilter queue} (\texttt{NFQUEUE}, as OpenSnitch uses) holds a packet in the kernel pending a user-space verdict, which satisfies G1 and G5, but it acts on the first packet after the socket is created and attributes by scanning for the owning process after the fact, which is racy under load and has no notion of workload identity. An \emph{LSM hook} (\texttt{security\_socket\_connect} via BPF-LSM) is synchronous and pre-packet, but its verdict is a local allow/deny: it cannot sleep to wait for an external verifier, so it cannot host a held, remotely-decided connection (G3, G4). A \emph{transparent redirect} (TPROXY, or NAT \texttt{REDIRECT}) delivers the connection to a local userspace process that can then take as long as it likes, which is what we need, but plain redirect loses the original destination and the identity unless they are recovered separately.

The \texttt{cgroup/connect4} hook combines the properties none of these has alone. It is synchronous and pre-packet (G1): it runs inside \texttt{connect(2)} before a route is allocated. It can \emph{rewrite} the destination, so it redirects the socket to a userspace interceptor that is free to wait for an external verdict (G3, G4, G5). It runs in process context, so it reads the cgroup id and pid directly from the kernel (G2). And it is attached per cgroup, so it covers every process in the pod --- the agent, its children, its tool servers --- without anything in the agent's address space (G6). The one thing it does not give us for free is the original destination on the interceptor side, because after the rewrite the kernel only knows the interceptor's address; we recover it from the socket cookie, which is the hinge of the whole design and the subject of Section~\ref{sec:attrib}.

\subsection{The hold point}
\label{sec:holdpoint}
\label{sec:hold}
A \texttt{BPF\_PROG\_TYPE\_CGROUP\_SOCK\_ADDR} program attached at \texttt{BPF\_CGROUP\_INET4\_CONNECT} runs inside \texttt{connect(2)} for every process in the cgroup, before the kernel allocates a route or sends a SYN. For TCP it does three things: it reads the socket cookie (\texttt{bpf\_get\_socket\_cookie}), which the kernel assigns lazily and guarantees unique for the life of the socket; it stores \{original address and port, \texttt{bpf\_get\_current\_cgroup\_id()}, \texttt{bpf\_get\_current\_pid\_tgid()}, time\} in an LRU hash map keyed by that cookie; and it overwrites \texttt{user\_ip4}/\texttt{user\_port} with the interceptor's address. The agent's socket then connects to the interceptor; from the agent's side \texttt{connect()} simply succeeds, with no indication that its destination was changed. The program is 98 instructions including the UDP rule and the per-rule counters, has no backward jumps, passes the verifier trivially, and runs once per connection. It is attached per pod cgroup with \texttt{BPF\_F\_ALLOW\_MULTI}, so several pods and other cgroup programs coexist, and the interceptor's and verifier's own connections --- which originate outside the supervised cgroups --- are never rewritten, so there is no redirection loop.

Two companion programs share the same maps. A \texttt{cgroup/sendmsg4} program applies the UDP policy of Section~\ref{sec:udp} to unconnected datagrams. A \texttt{sockops} program, invoked when a socket reaches \texttt{ACTIVE\_ESTABLISHED}, writes a second index, \{local address, local port\} $\rightarrow$ cookie, into a \texttt{peer\_map}. This index is what lets the interceptor recover the cookie of a connection it accepts from a pod in a \emph{different network namespace} (Section~\ref{sec:attrib}); without it, an \texttt{inet\_diag} lookup from the node's namespace would not see the pod's socket.

\paragraph{iptables fallback} Where eBPF cannot be loaded --- an older kernel, a managed node without \texttt{CONFIG\_BPF\_LSM}, a locked-down runtime --- an iptables rule, \texttt{-t nat -A OUTPUT -p tcp -m owner --uid-owner \$AGENT -j REDIRECT --to-ports \$INTERCEPTOR}, delivers the same socket to the interceptor through conntrack NAT, and the interceptor recovers the original destination with \texttt{SO\_ORIGINAL\_DST}. The rule keys on the agent's uid, so the interceptor must run as a different uid, exactly as service-mesh sidecars do to avoid intercepting their own egress. This path loses the kernel-supplied pid and cgroup id (it recovers them by scanning \texttt{/proc}) and the UDP pinning, and it assumes the agent runs unprivileged: an agent that is root inside its container could \texttt{setuid} a child to another uid and step out of the \texttt{--uid-owner} match, whereas the cgroup-keyed eBPF hook follows the child regardless. It needs no eBPF and is otherwise equivalent. Both paths pass the full adversarial suite (Section~\ref{sec:eval}).

\subsection{Attribution}
\label{sec:attrib}
When the interceptor \texttt{accept()}s the redirected connection, it must answer three questions about a socket it did not create: which kernel socket is this, which process and cgroup opened it, and which pod identity does that cgroup belong to. The three answers are joined through the socket cookie.

\paragraph{Kernel object} The interceptor recovers the cookie of the agent's socket in one of two ways. In the common case the \texttt{sockops} program has already written \{local address, local port\} $\rightarrow$ cookie into \texttt{peer\_map}, and the interceptor looks it up by the peer address of the accepted connection; this works across network namespaces, which matters because each pod has its own. As a fallback (and on the iptables path) it issues a \texttt{NETLINK\_SOCK\_DIAG} (\texttt{inet\_diag}) query for the exact 4-tuple, which returns the socket's cookie, inode and uid. Either way the cookie it obtains is \emph{the same value} the \texttt{connect4} program keyed its \texttt{escrow\_map} entry on, so the interceptor can read back the original destination, cgroup id and pid the kernel recorded at \texttt{connect()} time. No secret is shared between the agent and the interceptor for this to work; the join is on a kernel-assigned identifier the agent cannot forge.

\paragraph{Process} With the eBPF entry the pid and cgroup id are known directly. Otherwise the owning process is found by scanning \texttt{/proc/*/fd} for the symlink \texttt{socket:[inode]}. From \texttt{/proc/<pid>} the interceptor reads the cgroup path, the executable path, its SHA-256, and \texttt{comm}. The digest is taken by reading through \texttt{/proc/<pid>/exe}, which follows the kernel's reference to the actual inode, so an adversary that copies or renames its interpreter gets a new inode and the real digest of whatever it ran --- the renamed-binary bypass (T11) changes the path but not the digest or the cgroup.

\paragraph{Platform identity} A registry maps cgroup-path prefixes to pod identities. On Kubernetes the kubelet names each pod's cgroup after the pod uid, so the mapping is a longest-prefix lookup, and the node agent fills it from the API server for the pods it has labelled supervised, refreshing on an interval so pods that start later are covered. The identity carried forward is \{pod uid, namespace, name, service account, SPIFFE id\}; the SPIFFE id is the key the verifier's policy is written against.

The attribution record that results, and travels to the verifier, is in Table~\ref{tab:record}. Every field below the line is filled by the kernel or the node agent, none by the pod.

\begin{table}[t]
\caption{The attribution record for one held connection. Source column: which trusted component supplied each field.}
\label{tab:record}
\centering
\scriptsize
\resizebox{\columnwidth}{!}{%
\begin{tabular}{@{}lll@{}}
\toprule
Field & Meaning & Source \\
\midrule
\texttt{cookie} & kernel socket identifier & sockops / inet\_diag \\
\texttt{inode, uid} & socket inode, owner uid & inet\_diag \\
\texttt{pid, tgid} & owning task & eBPF map / \texttt{/proc} \\
\texttt{cgroup\_id, cgroup} & kernel cgroup id and path & eBPF map / \texttt{/proc} \\
\texttt{exe, exe\_sha256} & executable path and digest & \texttt{/proc/<pid>/exe} \\
\texttt{comm} & process name & \texttt{/proc} \\
\texttt{workload} & pod uid, ns, SA, SPIFFE id & registry \\
\texttt{dst} & original ip, port; redirect source & eBPF map / \texttt{SO\_ORIG\_DST} \\
\texttt{wire} & http/tls/opaque; host/SNI/path; len; hash & interceptor \\
\bottomrule
\end{tabular}}
\end{table}

\subsection{Parking the first bytes}
The interceptor reads up to 4\,KiB for at most 150\,ms. An HTTP/1.x request line and headers yield method, host and path; a TLS ClientHello yields the SNI; anything else is opaque. Nothing is forwarded: the bytes stay with the connection and go to the destination only after release, byte for byte. Their hash is part of what the verdict binds to, so an approval for one request cannot be spent on another written on the same socket.

\subsection{The verifier}
\label{sec:verifier}
The verifier is a service with its own signing key, running in its own pod. It accepts an escrow record authenticated with a message-authentication key (HMAC-SHA256) that is mounted only into the interceptor container, so a pod that somehow reached the verifier's port could not submit a record it would accept. It evaluates a policy keyed on the workload's SPIFFE id and returns a signed verdict
\begin{multline*}
v = \mathrm{Sign}_{\mathrm{verifier}}\big(\text{escrow\_id},\ \text{cookie},\ H(\text{dst},\ \text{first bytes}),\\
\text{ALLOW}\,|\,\text{DENY},\ t_{\mathrm{issued}},\ t_{\mathrm{expires}},\ \text{nonce}\big).
\end{multline*}

\paragraph{Policy} A policy is a set of per-workload rule lists plus a default. Each rule matches on any of: destination by name (SNI or HTTP Host, with glob), by Kubernetes service (a cluster-IP$\rightarrow$\texttt{namespace/name} map the verifier keeps from the API server), or by address; destination port; HTTP method and path prefix; and whether layer-7 metadata is required at all. The first matching rule decides \texttt{allow}, \texttt{deny} or \texttt{hold}. A per-workload token-bucket \emph{budget} caps the rate of allowed connections, so a workload that is compromised into a burst of approved-looking calls is throttled even where each call matches a rule. A rule set is deliberately small and declarative; the point of the paper is the enforcement path, and the policy function is a single call that any richer engine --- an OPA/Rego evaluator, a learned classifier, a human queue --- can replace without touching the kernel or the interceptor. Section~\ref{sec:verifier} ships the rule engine as a default, not as the contribution.

\paragraph{Service identity of destinations} Because pods reach services by cluster IP, a rule written against \texttt{lab/internal-api} must be matched against the cluster IP the connection actually used. The verifier maintains the reverse map from the API server and refreshes it, so a destination's \emph{service} identity, not just its IP, is available to the policy and recorded in the receipt.

\paragraph{Self-protection} A destination equal to the verifier's own address (or that of its service) is always denied with reason \texttt{destination\_is\_verifier}. Combined with the interceptor-only record key, this closes G3 from both sides: the agent cannot reach the verifier (its connection to it is escrowed and refused), and could not be understood by it if it did.

\subsection{UDP and DNS pinning}
\label{sec:udp}
TCP connections are escrowed; UDP is \emph{pinned}. An agent that cannot exfiltrate over TCP could otherwise leak over DNS --- encoding data in query names to a resolver it controls --- or over any other UDP service. The \texttt{connect4} and \texttt{sendmsg4} programs therefore apply a narrow rule to UDP: a datagram to the configured cluster resolver (\texttt{dns\_ip:53}) is allowed; every other UDP destination is denied in-kernel with \texttt{EPERM}. This is a pin rather than an escrow --- the datagram is not held for a verdict --- because DNS is latency-critical and the cluster resolver is a fixed, trusted endpoint; holding every query would add a round trip to every name lookup for no security gain. The effect is that name resolution works normally through the cluster's own resolver, and a query aimed anywhere else, or a UDP tunnel to an arbitrary host, never leaves the node. Per-rule kernel counters record how many UDP datagrams were allowed and denied, which the evaluation reports. The same approach extends to other UDP services (NTP, QUIC) by widening the pinned set; escrowing UDP for a verdict, and the IPv6 hooks, are left to future work (Section~\ref{sec:limits}).

\subsection{Gating on content without terminating TLS}
\label{sec:content}
Everything so far decides on \emph{where} the agent connects --- destination, port, and the SNI visible in a TLS ClientHello. That is the right control for the dominant agent-egress threats (exfiltration to an unauthorized destination, command-and-control, calling a forbidden service), and encryption hides none of it: the destination address is the one field the kernel must have in the clear to route, and escrow sits exactly where it is. What encryption does hide is the request \emph{content} --- the body of a request to a destination the policy permits. To gate on that, one normally terminates TLS at a proxy with a trusted certificate authority in the workload, which holds a key, sees all plaintext, and breaks the end-to-end property. We avoid that.

\paragraph{Reading the plaintext} The plaintext is available at the agent's own endpoint, before encryption. An eBPF uprobe attached to the write function of the agent's TLS library is invoked with the plaintext buffer as an argument; it copies up to one TLS record (16\,KiB) of every write, in order, into a kernel FIFO, each event carrying the writing task, the full length of the write, and the TLS object it was made on (the \texttt{SSL*} or \texttt{*tls.Conn} handle, the function's first argument). A second pair of probes records which socket each handle writes to: exactly, by a probe on \texttt{SSL\_set\_fd} for OpenSSL, or by the socket write that immediately follows a TLS write on the same task (\texttt{syscall.write} for Go, \texttt{write} in libc as a fallback). The interceptor resolves that handle to a file descriptor and, through \texttt{/proc}, to the socket's local port, which is the port it accepted the connection from, so each captured write is matched to its connection by socket identity rather than by arrival order. Where that function lives, and which registers carry the buffer, depend on the runtime, and nothing else does. For OpenSSL the symbols are \texttt{SSL\_write} and \texttt{SSL\_write\_ex} with the buffer and length in \texttt{rsi} and \texttt{rdx} (System~V ABI); the probe is placed at the symbol's file offset in \texttt{libssl.so} when the library is shared, or in the agent's own executable when it is statically linked, which is also the shape of a BoringSSL build. For Go, \texttt{crypto/tls} is compiled into every binary; the symbol is \texttt{crypto/tls.(*Conn).Write} and the Go register ABI passes the slice pointer and length in \texttt{rbx} and \texttt{rcx}, so a second 52-instruction program reads those two registers and shares the capture path. The loader inspects the agent's ELF to choose the program and the symbol (Go symbol present $\rightarrow$ Go; \texttt{SSL\_write} present $\rightarrow$ static; otherwise shared \texttt{libssl}). The TLS session is untouched --- the handshake completes end-to-end to the real server, which presents its real certificate, and the agent validates it against the hostname as usual. The enforcer never possesses a key and never decrypts the wire; it observes the plaintext inside the agent's address space, through the kernel. This capture technique is itself known from observability systems~\cite{zheng2025agentsight}; the contribution is to \emph{act} on it.

\paragraph{Holding on the plaintext} The interceptor already relays the agent's encrypted bytes without a key. For a TLS connection it forwards the handshake untouched and then holds \emph{every} outgoing application-data record until the write that produced it has been scored. The simplest form of the gate --- hold only the first application-data record, then splice --- is the one we built first, and it is wrong in a way worth stating: it is blind to everything written after that record, so a benign first request on an HTTP keep-alive connection opens the door for a second, hostile request on the same socket, the tail of a body longer than one record is never seen, and under HTTP/2 the first record is the connection preface, not a request at all. Section~\ref{sec:evalcontent} demonstrates that exploit before showing it closed. The \emph{stream gate} drains the FIFO, binds the connection to the TLS handle of its first write and thereafter consumes only writes on that handle, so concurrent TLS sessions in one process cannot cross; a write of $n$ plaintext bytes becomes one or more records, and the relay calibrates the per-record ciphertext overhead from the first record and releases records until $n$ bytes are accounted for, then treats the next record as the start of a new write and fetches its plaintext. A content model scores each write; on \texttt{allow} its records are released, on \texttt{deny} the connection is reset and the held record never leaves the node. If no plaintext arrives for a held record within the timeout the gate denies, so an agent whose TLS the probe cannot see loses egress rather than gaining an unobserved channel. The verdict is bound to the connection as in Section~\ref{sec:gate}; this is the same single-use, identity-bound gate applied one layer up, to each write instead of to the destination. The cost is that a content-gated connection stays in the userspace relay for its lifetime and pays one model call per write. Under TLS~1.2 application-data records (type \texttt{0x17}) are distinguishable from the handshake on the wire; under TLS~1.3 the outer type is uniformly \texttt{0x17} after the ServerHello, so the gate holds every post-ServerHello record and uses the uprobe events, which exist only for application writes, to decide what to score --- the enforcer still decrypting nothing.

\paragraph{Scope} The uprobe is per-library. Shared OpenSSL, statically linked OpenSSL (and by the same mechanism BoringSSL, which keeps the \texttt{SSL\_write} signature) and Go's \texttt{crypto/tls} are covered and evaluated (Section~\ref{sec:evalcontent}); Rustls has a different write shape and needs its own probe, and a stripped binary carries no symbol, so its offset must be supplied out of band. The content model is a deterministic logistic-regression placeholder over request features; like the connection policy, it is replaceable and is not the contribution. Two limits remain in the prototype. Up to one TLS record (16\,KiB) of each write is captured, so a request whose body spans several records is seen only up to its first record and a model that scores whole multi-record bodies needs the capture chunked across them. And under HTTP/2 every frame is scored at the byte level, but HEADERS frames are HPACK-compressed, so a content model needs a per-connection HPACK decoder to read them; DATA frames are plain. The join from a captured write to its connection is by the write's socket, resolved through \texttt{/proc}, so concurrent TLS sessions in one process are disambiguated whatever their interleaving (Section~\ref{sec:evalcontent} runs two interleaved sessions and confirms an exfiltration write stays on its own connection); the resolution races the kernel's fd binding by a few microseconds, during which the gate falls back to the handle of the connection's first write. More fundamentally, the content gate's adversary is weaker than the connection gate's. Section~\ref{sec:problem} grants the adversary full code control; against that adversary the connection gate holds, because it is in the kernel and keyed on the cgroup, but the content gate does not: an agent that ships its own TLS, strips its symbols, or inlines encryption gives the probe nothing to see. The gate then fails closed --- the agent loses TLS egress rather than gaining a channel --- but a decoy write through the probed function while the real payload goes another way, or a race on the buffer between the probe's copy at function entry and the library's encryption, is available to such an adversary and is inherent to observing before encryption inside an address space the adversary owns. The content gate is therefore a control against hostile \emph{input} to an honest agent, the prompt-injection setting of the introduction, not against hostile code; the connection gate is the control against hostile code. What is new is the combination --- see the plaintext via a uprobe, gate the connection on it with a single-use identity-bound verdict, and keep the connection end-to-end with no key at the enforcer --- which removes the usual ``inspect or stay end-to-end, choose one'' tradeoff between a terminating proxy and a pure observer.

\subsection{The release gate}
\label{sec:gate}
The interceptor releases a held connection only if all of the following hold: the escrow id is still live (not decided, not retired by the deadline); the signature verifies under the verifier's key; the verdict names this escrow id, this socket cookie and this destination hash; the verdict has not expired; the escrow id has not been consumed; the verdict is ALLOW. On success the escrow id is consumed and the interceptor connects to the destination, forwards the parked bytes and splices the two sockets. Any other outcome, including a transport error toward the verifier, sets \texttt{SO\_LINGER} to zero and closes, so the agent sees a reset. Because the interceptor only calls \texttt{connect()} after the gate passes, a refused connection sends nothing to the destination (G1).

\subsection{Hold and release with a deadline}
\label{sec:hold-release}
\begin{figure*}[t]
\centering
\includegraphics[width=0.9\textwidth]{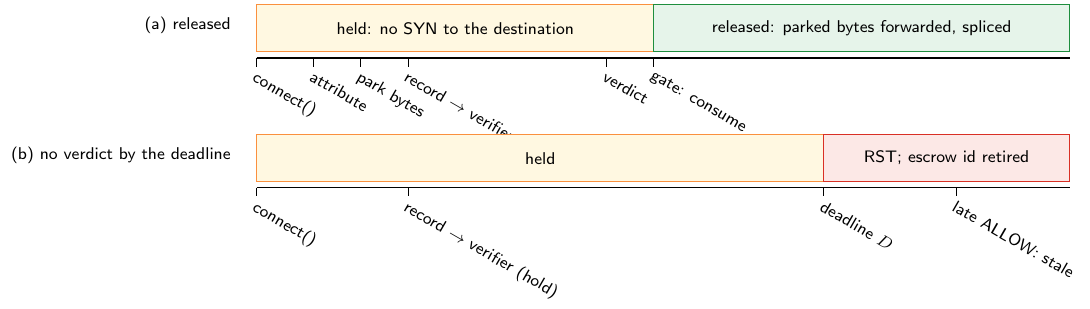}
\caption{Timeline of one escrowed connection. (a) Released: the destination sees the first packet only after the gate consumed the verdict. (b) No verdict by the deadline $D$: reset, escrow id retired, a late ALLOW is stale.}
\label{fig:timeline}
\end{figure*}

A \texttt{hold} rule parks the decision at the verifier until an operator posts a release or the verifier's hold timeout passes. The interceptor's deadline $D$ is independent and shorter. When $D$ passes the connection is reset and the escrow id is retired; a verdict released afterwards for that id fails the first condition of the gate (``stale'') and opens nothing (Fig.~\ref{fig:timeline}). An approval therefore cannot act on a connection the agent attempted earlier, and the agent cannot keep a connection pending indefinitely.

\subsection{Receipts}
Each decision becomes a receipt \{escrow id, node, cookie, inode, pid, cgroup, executable digest, workload, destination, wire summary, outcome, reason, timings\} appended to a hash chain at the interceptor; the verifier keeps its own chain of \{escrow id, workload, decision, reason, decision time\}. A changed receipt breaks the chain at that entry.

\subsection{Deployment on Kubernetes}
\label{sec:deploy}
A DaemonSet runs the node agent with host PID and host network, the host cgroup tree mounted, and the capabilities needed to load and attach the programs. It watches the API server for pods on its node labelled \texttt{escrow.io/supervised=true}, resolves each pod uid to its cgroup directory under the node's \texttt{kubepods} hierarchy, attaches the three programs there, and writes the registry; the watch interval picks up pods that start later. The interceptor listens on the node address reachable from pods (the CNI bridge), so a redirected connection from any pod on the node arrives at it; because it runs on the host network it can in turn reach cluster-IP services (through kube-proxy) and the verifier. The verifier is a Deployment in its own namespace, fronted by a Service, with a NetworkPolicy admitting only node addresses on its port. The supervised pod needs nothing but the label and its service account (G6): no sidecar, no init container, no changed image. For clusters where eBPF cannot be loaded, a per-pod sidecar variant uses an \texttt{iptables} init container and a shared process namespace so a sidecar interceptor can attribute through \texttt{/proc}. Section~\ref{sec:k8s} reports a run of this deployment against real pods on a single-node cluster.

For the interceptor to gate a short-lived agent's \emph{first} connection, the programs must be attached to the pod's cgroup before the agent runs. In the steady state of a long-lived agent the watch loop handles this; for the general case the node agent can attach on the pod-creation event before the container's entrypoint has made a call, and a pod that briefly predates attachment is covered from its next connection onward. Long-running agent workloads --- the target of this work --- are attached once and supervised for their lifetime.

\section{Implementation}
\label{sec:impl}
The prototype is Python 3 (standard library; \texttt{pyelftools} to load the ELF object; \texttt{cryptography} optionally for native Ed25519, with a pure-Python fallback) and about 250 lines of restricted C compiled with \texttt{clang -target bpf} into five programs: \texttt{connect4} (98 instructions), \texttt{sendmsg4} (50), \texttt{sockops} (22) and the two TLS-write uprobes (52 each). The loader uses \texttt{bpf(2)} directly: it creates the two maps, patches the \texttt{ld\_imm64} map relocations, loads the program as \texttt{CGROUP\_SOCK\_ADDR}/\texttt{INET4\_CONNECT} and attaches it to a cgroup directory with \texttt{BPF\_F\_ALLOW\_MULTI}. The interceptor is an \texttt{asyncio} server; attribution runs in a thread. The verifier is a threaded HTTP server with \texttt{/v1/escrow}, \texttt{/v1/release} and \texttt{/v1/pending}. Records are canonical JSON; verdicts and receipts are Ed25519 over a tagged canonical encoding. The whole system, both hold points, the harness, the suite and the benchmark are about 3\,800 lines, of which the system itself (package and eBPF programs) is about 2\,150.

The test harness creates one cgroup per pod under cgroup v2, runs each agent as a separate process under an unprivileged uid inside that cgroup, and uses loopback echo servers that record every accepted connection and its first bytes. Agents are ordinary programs: the suite uses a small Python client, and the end-to-end script uses an unmodified \texttt{curl}. The cross-engine primitives (canonical JSON, Ed25519, the hash-chained audit log) are reused from our earlier gateway~\cite{zkproofgateway}, so the new code is the kernel programs, the loader, the attribution join, the interceptor state machine and the verifier.

\section{Security analysis}
\label{sec:security}
We argue each goal against the adversary of Section~\ref{sec:problem}, and point to the adversarial check (Section~\ref{sec:eval}, Table~\ref{tab:suite}) that exercises it.

\paragraph{G1, pre-packet hold} The interceptor calls \texttt{connect()} toward the destination only after the release gate returns true. On the eBPF path the agent's own \texttt{connect()} never carried the destination address past the hook --- the kernel rewrote it in place --- so for a refused connection no SYN is ever formed for the real destination. The harness confirms the destination accepts nothing (T1, T7); the code path is what makes the guarantee, not the observation. The one caveat is that the destination's address appeared in the kernel's \texttt{escrow\_map}, never on the wire.

\paragraph{G2, attribution, under evasion} Attribution binds to the cgroup and the executable digest, not the path or the process name. A child process (T11) is in the same cgroup and is attributed to the same pod; a renamed or copied interpreter (T11) yields a new inode and the true digest of what ran. The socket cookie that joins the kernel and interceptor views is assigned by the kernel and cannot be chosen by the agent, so the agent cannot present one connection's identity for another's.

\paragraph{G3, isolated verifier} The verifier is a destination like any other, so an agent's attempt to reach it is itself escrowed and refused by the \texttt{destination\_is\_verifier} rule (T6); only the interceptor's record reaches the verifier at all. And the record carries an HMAC under a key mounted only in the interceptor, so even a pod that reached the verifier's port could not submit a record it would accept. The agent can thus neither request its own approval nor answer as the interceptor.

\paragraph{G4, single-use socket-bound release} The gate accepts a verdict only if it is signed by the verifier (forgery under another key fails, T5), names this escrow id (a verdict for another connection fails), names this socket cookie (rebinding to another socket fails, T5), names the hash of this destination and these first bytes (rebinding to another destination or request fails, T5), has not expired (T5), and has not been consumed (replay of an accepted verdict fails, T5). Consumption is atomic: the escrow id moves to a consumed set under the interceptor's lock, so two concurrent deliveries of the same verdict cannot both release. This is the property that closes TOCTOU: an approval is spent on exactly the one socket it names.

\paragraph{G5, fail closed} If no verdict arrives within the deadline $D$, or the verifier is unreachable or returns a transport error, the gate's first condition fails and the connection is reset (T3, T12). The deadline is enforced at the interceptor independently of the verifier's own hold timeout, so a slow verifier cannot hold a connection open past $D$. A verdict that arrives after $D$ finds the escrow id retired and is rejected as stale (T4), so delaying or replaying a verdict cannot reach a fail-open state; the system's failure direction is always closed.

\paragraph{G6, no agent change} Nothing in the enforcement lives in the agent's address space. The hook is in the kernel, attached to the cgroup, and covers the agent, its children and its tool servers regardless of language or runtime. The bypass comparison (Section~\ref{sec:k8s}) makes this concrete: a control that lives in the agent's environment (an \texttt{HTTP\_PROXY} setting) is defeated by a raw socket, while the kernel hook holds it.

\paragraph{Residual risk} The analysis assumes the TCB of Section~\ref{sec:problem}. An adversary with root on the node can detach the programs or read the keys; escrow does not defend the node against itself, and a production deployment would pair it with attestation of the loaded program set (which our earlier gateway~\cite{zkproofgateway} provides and which is left as integration). Data-only attacks that reuse an already-released connection are out of scope for the connection gate, which controls the opening of a connection and the first bytes on it; the content gate's stream mode extends the check to every write on a content-gated connection (Section~\ref{sec:content}). UDP other than to the pinned resolver is denied, not escrowed, so a covert channel inside allowed DNS traffic is not addressed.

\section{Evaluation}
\label{sec:eval}
The adversarial suite and the microbenchmark (Sections~\ref{sec:suite}--\ref{sec:latency}) are from one logged run (\texttt{results/execution\_log.txt}, 19 Oct 2026) on a 2-vCPU virtualised Intel Xeon at 2.80\,GHz, Linux 6.18, cgroup v2, Python 3.13, clang 18; everything there is on loopback, so network latency is absent and the figures isolate the cost of the mechanism. The cluster run (Section~\ref{sec:k8s}) and the content-gate runs (Section~\ref{sec:evalcontent}) are logged separately on the same host, as noted in each.

\subsection{Adversarial suite}
\label{sec:suite}
Table~\ref{tab:suite} lists the checks. Each runs the same way: an agent process in a pod cgroup does something, the harness records what the destination accepted, and the receipt is inspected. All checks pass on both hold points: 36 on the eBPF path, 33 on iptables (the three UDP-pinning checks of T14 need the kernel path and are recorded as not applicable there).

\begin{table}[t]
\caption{Adversarial suite. ``Dst'' is what the destination accepted. 36 checks on the eBPF path, 33 of which apply on iptables; all pass.}
\label{tab:suite}
\centering
\scriptsize
\begin{tabularx}{\columnwidth}{@{}lXl@{}}
\toprule
\# & Scenario and expected behaviour & Goal \\
\midrule
T0 & Allowed destination: released; dst accepts exactly one connection carrying the parked bytes; receipt names the agent's pid, pod cgroup, SPIFFE id, cookie, executable digest, \texttt{source=bpf\_connect4} & G2 \\
T1 & Denied destination: agent gets a reset; dst accepts nothing & G1 \\
T2 & Same uid outside any pod cgroup: refused (\texttt{unknown\_workload}) even for an allowed destination (iptables path) & G2 \\
T3 & Hold rule, no release: reset at $D=1000$\,ms (measured 1001\,ms), dst untouched & G5 \\
T4 & Correctly signed ALLOW for that escrow id delivered after $D$: refused (stale); the verifier's late release opened nothing & G5 \\
T5 & Against a live escrow: verdict signed by another key; verdict for another cookie; for another destination hash; expired; valid verdict accepted once then refused on replay & G4 \\
T6 & Agent connects to the verifier's address: escrowed, refused (\texttt{destination\_is\_verifier}); only the interceptor's record reached the verifier & G3 \\
T7 & TLS: ClientHello with allowed SNI released and delivered byte for byte; other SNI reset, dst untouched & G1 \\
T8 & HTTP: \texttt{GET /v1/} released; \texttt{POST /v1/} and \texttt{GET /admin} refused & -- \\
T9 & Budget of 2 per minute for pod-b: third connection refused & -- \\
T10 & Hold then operator allow within $D$: released; hold then operator deny: reset & G5 \\
T11 & Connect from a child process: still held and refused; copied and renamed interpreter: still attributed to the pod by cgroup, same executable digest & G2, G6 \\
T12 & Verifier stopped: allowed destination refused (\texttt{DENIED\_VERIFIER\_UNREACHABLE}) & G5 \\
T13 & Both receipt chains verify; one edited receipt breaks the chain at that entry & -- \\
T14 & UDP (eBPF path): queries to a public resolver and UDP exfil refused in-kernel; the cluster resolver allowed; DNS over TCP to a public resolver escrowed and refused & G1 \\
\bottomrule
\end{tabularx}
\end{table}

Two details matter for G1. The harness's ``destination accepted nothing'' is an accept-level observation; the stronger statement follows from the code path: the interceptor calls \texttt{connect()} toward the destination only after \texttt{accept\_verdict} returns true, and with the eBPF hold point the agent's own \texttt{connect()} never carried the destination address past the hook. For T3, the agent's \texttt{connect()} returned with a reset 1.07\,s after it began, consistent with the 1\,000\,ms deadline plus process start.

\subsection{Latency}
\label{sec:latency}
Table~\ref{tab:latency} breaks a released connection into its stages, from the interceptor's receipts and the agent's own clock, over 200 sequential connections, against an unsupervised baseline to the same echo server.

\begin{table}[t]
\caption{Latency per released connection, median / p95 in ms, $N=200$ sequential; then 16 concurrent agent processes with 12 connections each. Native Ed25519 via \texttt{cryptography}; pure Python otherwise.}
\label{tab:latency}
\centering
\scriptsize
\setlength{\tabcolsep}{3pt}
\resizebox{\columnwidth}{!}{\input{figures/tab_latency_body.tex}}
\end{table}

The hold costs 4.3\,ms median on the eBPF path (6.8\,ms p95) and 5.8\,ms on the iptables path, against a 0.7\,ms baseline. Policy evaluation itself is 24\,$\mu$s. The rest is one fresh HTTP/1.1 round trip to the verifier including its signature (2.2\,ms), attribution (0.9\,ms on the eBPF path, where the kernel already named the pid; 2.0\,ms on the iptables path, which scans \texttt{/proc}), the gate's signature verification (0.3\,ms native, 3.3\,ms pure Python) and the outbound connect (0.4\,ms). With pure-Python Ed25519 the hold is 9.3\,ms: the two signature operations account for most of the difference, which puts an upper bound on what a native verifier and a persistent connection would recover. Under 16 concurrent agents on two vCPUs the median hold rises to 32\,ms (48\,ms p95) at 126 connections per second; the throughput figure varies by run on a shared two-vCPU host, and the queueing is in the single-process Python interceptor and the threaded verifier, which share the CPUs with the sixteen agents, not in the kernel program.

For an agent whose tool calls take hundreds of milliseconds to seconds each, a 5\,ms hold per connection is below the noise of the call itself. The deadline, not the hold, is the operative cost: a hold rule that waits for an operator pays the full deadline when nobody answers.

\subsection{End to end with an unmodified client}
\texttt{experiments/cli\_smoke.sh} starts the verifier and the interceptor from the command line, attaches the program to a fresh cgroup, moves the shell into it, and runs \texttt{curl}. The allowed port answers (receipt: \texttt{RELEASED}, \texttt{comm=curl}, \texttt{source=bpf\_connect4}); \texttt{curl} against the verifier's port, and the same attempt from a child shell, are reset (\texttt{destination\_is\_verifier}).

\subsection{On a Kubernetes cluster}
\label{sec:k8s}
\begin{figure*}[t]
\centering
\includegraphics[width=0.96\textwidth]{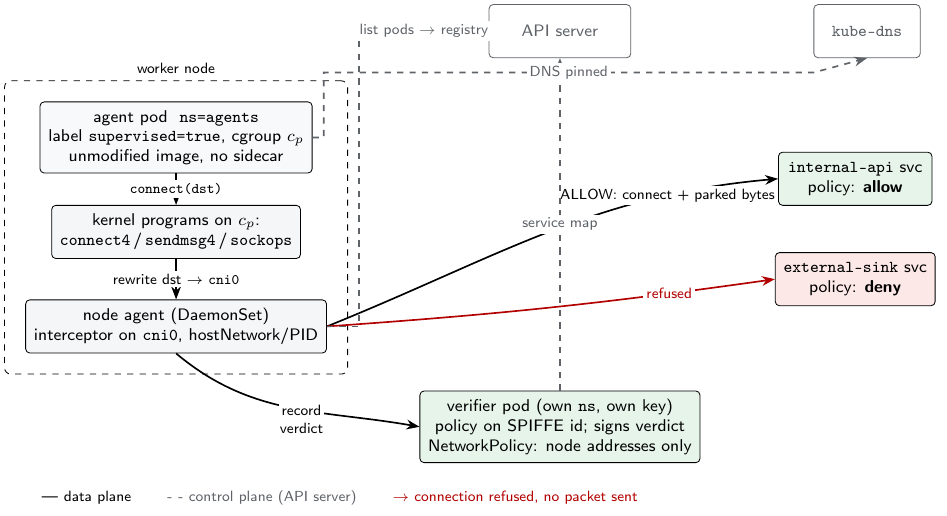}
\caption{Deployment on Kubernetes. A supervised pod's \texttt{connect()} is held in the kernel and redirected over \texttt{cni0} to the node-agent interceptor, which asks the verifier and, on ALLOW, connects to the destination Service through kube-proxy. The node agent builds its pod registry and the verifier its cluster-IP service map from the API server (control plane, dashed); DNS is pinned to the cluster resolver. The agent pod carries only a label; it runs no sidecar.}
\label{fig:cluster}
\end{figure*}

Fig.~\ref{fig:cluster} shows the deployment. We ran the system on a single-node Kubernetes cluster (k3s v1.34.1, upstream Kubernetes, cgroup v2) with the node agent (verifier, interceptor and eBPF programs) as a host process against real pods, exactly as the DaemonSet in \texttt{deploy/k8s} deploys it. The cluster has two destination Services: \texttt{internal-api}, which the agent's workload identity may reach, and \texttt{external-sink}, which it may not. Agent pods are unmodified images running a plain \texttt{urllib} client; the node agent resolves each supervised pod's SPIFFE identity from its cgroup and attaches the programs to the pod cgroup before the pod makes any call. All reported numbers are from one logged run (\texttt{results/RESULTS\_K8S.md}).

\textbf{Enforcement.} An agent that issues GET and POST to \texttt{internal-api} and to \texttt{external-sink}, and spawns a child process that also posts to \texttt{external-sink}, reaches every destination when unsupervised: \texttt{external-sink} records the posted body. Under escrow, the two \texttt{internal-api} calls are released (receipts \texttt{rule[0]}, \texttt{rule[1]}, attributed to the pod's SPIFFE id, cgroup, pid and the resolved Host name) and the three \texttt{external-sink} connections, including the child's, are reset before a packet reaches the Service; \texttt{external-sink} records nothing. The kernel counters show every TCP connect held and every DNS query pinned to the cluster resolver (none leaked to an external resolver), as the \texttt{sendmsg4}/\texttt{connect4} UDP rule enforces.

\textbf{Bypass comparison.} We compared escrow with a sidecar-style egress proxy, the strict case of a per-request proxy control: the agent pod's \texttt{HTTP\_PROXY} points at a deny-all proxy. The agent then makes one HTTP call to \texttt{external-sink} (which consults the proxy) and one raw-socket call to \texttt{external-sink}'s address (which does not). Under the proxy the HTTP call is refused but the raw socket reaches the Service, and the data leaves the pod; under escrow both are held at \texttt{connect()} and nothing leaves (Table~\ref{tab:bypass}). This is the gap the introduction describes: a control the workload can route around versus one at the operating-system boundary it cannot.

\begin{table}[t]
\caption{Bypass comparison on the cluster. Both controls aim to keep the agent off \texttt{external-sink}. ``Received'' is the number of posts \texttt{external-sink} actually stored.}
\label{tab:bypass}
\centering
\scriptsize
\begin{tabular}{lccc}
\toprule
Control & HTTP-client call & raw-socket call & received \\
\midrule
Sidecar proxy (\texttt{HTTP\_PROXY}) & blocked (403) & \textbf{reached} & 1 \\
Connection escrow (kernel) & blocked (reset) & blocked (reset) & 0 \\
\bottomrule
\end{tabular}
\end{table}

\textbf{Scale.} Six supervised pods each opening 100 connections to \texttt{internal-api} (600 in all), through one interceptor on two shared vCPUs, were all held and released (600/600), at 59 connections per second, with a per-connection hold latency of 18\,ms median, 38\,ms p95 and 57\,ms p99. These are higher than the loopback microbenchmark of Section~\ref{sec:eval} because each connection now traverses the real pod network (pod~$\rightarrow$~\texttt{cni0}~$\rightarrow$~host interceptor~$\rightarrow$~ClusterIP via kube-proxy~$\rightarrow$~backend pod) and carries a per-connection DNS lookup, and because the interceptor, the verifier, the six agent pods and k3s share two CPUs. The kernel hook (one map update per connect) is not the bottleneck; the single-process Python interceptor is, and the limitations below say how it scales out.

\subsection{Content gate, without terminating TLS}
\label{sec:evalcontent}
We ran the content gate (\texttt{experiments/tls\_gate\_demo.py all both}) against a real HTTPS server with its own certificate, over TLS~1.2, with three agents that differ only in their TLS runtime: a Python agent using shared OpenSSL~3 (probe on \texttt{SSL\_write\_ex} in \texttt{libssl.so.3}), a C agent statically linked against \texttt{libssl} (probe on \texttt{SSL\_write} inside the agent's executable), and a Go agent using \texttt{crypto/tls} (probe on \texttt{crypto/tls.(*Conn).Write} inside the Go binary). The relay, the model and the gate are the same code in all three; only the probe placement differs. Each agent runs the exploit a first-record check cannot see: it opens \emph{one} TLS connection, validates the server's certificate, sends a benign \texttt{GET /v1/items} with \texttt{Connection: keep-alive}, reads the reply, and then sends \texttt{POST /transfer} carrying a card-like secret on the same connection. Each agent ran twice, once against the first-record gate and once against the stream gate (Table~\ref{tab:runtimes}).

Against the first-record gate the exploit works on all three runtimes: the \texttt{GET} is scored \texttt{allow} ($p=0.001$), the connection is spliced, and the \texttt{POST} rides the released connection to the server, which records the secret. Against the stream gate the \texttt{GET} is scored \texttt{allow} and released; the \texttt{POST} is the next write on the same handle, is scored \texttt{deny} ($p=0.999$, on features \{sensitive keyword, card-like number, sensitive path, mutating method\}), and the connection is reset with that record held --- the server records only \texttt{/v1/items}, and the agent sees a full keep-alive reply to its first request and a reset on its second. In all six runs the agent's certificate validation succeeded, proving the real server's certificate reached it unbroken, and the relay and the uprobe held no key and decrypted no wire bytes. A separate check (\texttt{experiments/tls\_gate\_multirecord.py}) sends a 40\,KiB benign write, which the library emits as three TLS records, followed by the exfiltration write: the stream gate scores the first write once, releases all three of its records, recognises the fourth record as a new write and denies it, confirming the record-to-write accounting. A further check (\texttt{experiments/tls\_gate\_concurrent.py}) opens two TLS connections from one process and interleaves their writes under a barrier, one connection sending only benign requests and the other an exfiltration request; the stream gate, binding each connection to its own socket, allows the benign connection to completion and denies the exfiltration write on the other, which never reaches the server --- a gate binding by arrival order would have risked crossing them. The Go runs matter most in practice: Go binaries are statically linked and carry their own TLS, so they are exactly the agents and tool servers a \texttt{libssl} probe would miss, and Go's \texttt{net/http} client negotiates HTTP/2, where the first record is never a request.

\begin{table}[t]
\centering
\caption{Content gate across TLS runtimes and gate forms. One keep-alive TLS connection per run: benign \texttt{GET}, then exfiltration \texttt{POST} on the same connection (\texttt{results/tls\_gate\_runtimes.json}). The agent verified the server certificate in every run; the relay held no key.}
\label{tab:runtimes}
\scriptsize
\begin{tabular}{@{}p{1.75cm}p{1.0cm}p{1.75cm}p{2.0cm}@{}}
\toprule
Runtime (probe in) & Gate & \texttt{POST} on same conn. & Server received \\
\midrule
OpenSSL shared (\texttt{libssl.so.3}) & first & not inspected & \texttt{GET} and \texttt{POST} \\
 & stream & deny $.999$, reset & \texttt{GET} only \\
OpenSSL static (agent binary) & first & not inspected & \texttt{GET} and \texttt{POST} \\
 & stream & deny $.999$, reset & \texttt{GET} only \\
Go \texttt{crypto/tls} (agent binary) & first & not inspected & \texttt{GET} and \texttt{POST} \\
 & stream & deny $.999$, reset & \texttt{GET} only \\
\bottomrule
\end{tabular}
\end{table}

This is the capability the introduction sets out, now holding for the whole connection and not only its first record: deciding on request content while the tunnel stays end-to-end, which a terminating proxy cannot do without a key and a pure observer cannot enforce.

\section{Related work}
\label{sec:related}
\textbf{Synchronous external authorization.} Envoy's \texttt{ext\_authz} filter~\cite{envoyextauthz} holds an HTTP request or a TCP connection for a check by an external service, with a timeout and a fail-closed option; Istio's \texttt{CUSTOM} authorization action~\cite{istioauthz} is built on it. These provide the hold, the external verdict and the deadline at a proxy, for traffic that passes through the proxy, with identity at the pod or mTLS-principal level. They do not see the process, do not bind the verdict to a kernel socket, and are bypassed by anything that does not traverse the sidecar. \textbf{Interactive host firewalls.} OpenSnitch~\cite{opensnitchfaq,lwnopensnitch} queues each new connection in NFQUEUE, identifies the process, asks the user, and applies a default after 15\,s; a ``once'' duration makes a verdict per connection. It is the closest non-agent prior art for a held, process-attributed connection, but has no workload identity, no separate verifier and no socket-bound token. \textbf{Quarantine until a later verdict} is old~\cite{US7080408B1}.

\textbf{Kernel enforcement for agents.} AgentSight~\cite{zheng2025agentsight} observes agents with eBPF (TLS uprobes correlated with syscalls) but does not enforce. ActPlane~\cite{zheng2026actplane} compiles information-flow rules into eBPF with block, kill and notify effects. mcpguard-dynamic~\cite{mcpguarddynamic} derives allow-lists from MCP tool schemas and enforces them with BPF LSM on \texttt{open}, \texttt{connect} and \texttt{execve}. ProcRoute~\cite{procroute2026} authorizes routes per process with \texttt{cgroup/connect4}, synchronously, from local policy. Tetragon~\cite{tetragonselectors} enforces in-kernel with override or kill. AgentKernel~\cite{zou2026agentkernel} argues for a mandatory enforcement boundary below the agent. In all of these a policy entry is standing: it authorizes any matching connection for as long as it is loaded, and nothing waits for an external verdict. AARM~\cite{errico2026aarm} specifies intercept-then-decide for agent actions, listing eBPF among possible architectures, without a socket-level mechanism. \textbf{Agent sandboxes.} Claude Code's sandbox~\cite{claudecodesandbox} routes egress through a proxy outside the sandbox and prompts the user for a new host, then allows the host for the session; NVIDIA OpenShell~\cite{openshell} forces egress through a CONNECT proxy under OPA, identifies the calling binary via \texttt{procfs}, and blocks unknown endpoints until an operator widens the policy. Both approve a destination, not a connection, and act on traffic that uses the proxy. Tran et al.~\cite{tran2026networking} argue for treating agent security as a networking problem with a per-agent sidecar under a separate control plane, which is the architectural position this paper implements at the kernel rather than at a proxy.

\textbf{Attribution building blocks.} Socket cookies are kernel-unique identifiers readable from eBPF~\cite{bpfsocketcookie} and, as we use here, from \texttt{inet\_diag}; cgroup-based container attribution with \texttt{connect} hooks was described by Meta~\cite{lpc2023metabpf}. SPIFFE/SPIRE resolves a process to a pod identity through the same cgroup path. We know of no prior system that keys an external verifier's single-use approval on the socket cookie.

Table~\ref{tab:related} places the closest systems against the five properties that together define connection escrow: (a) holding a connection pending an \emph{external} verdict rather than deciding locally; (b) attributing it to the process and workload identity; (c) enforcing before any packet leaves; (d) a deadline after which the held connection is denied; (e) binding each approval to a single socket, used once. Each property appears somewhere in the prior art; no prior system has all five. What distinguishes connection escrow is exactly that combination: hold at \texttt{connect()} inside the pod's cgroup, attribution that joins the kernel view to the pod identity, a verifier the agent cannot reach, and a release bound to one socket and one destination with a fail-closed deadline. Together they close the gap between approving an action and the connection that performs it --- the gap a standing allow-list or a bypassable proxy leaves open.

\begin{table}[t]
\caption{Closest systems vs.\ the five properties of connection escrow. \checkmark\ present, $\circ$ partial, blank absent. (a) external hold, (b) process/workload attribution, (c) pre-packet, (d) deadline$\rightarrow$deny, (e) single-use socket-bound.}
\label{tab:related}
\centering
\scriptsize
\begin{tabular}{@{}lccccc@{}}
\toprule
System & (a) & (b) & (c) & (d) & (e) \\
\midrule
Envoy \texttt{ext\_authz} / Istio~\cite{envoyextauthz,istioauthz} & \checkmark & $\circ$ & & \checkmark & $\circ$ \\
OpenSnitch~\cite{opensnitchfaq} & \checkmark & $\circ$ & \checkmark & \checkmark & $\circ$ \\
AgentSight~\cite{zheng2025agentsight} (observe only) & & \checkmark & & & \\
ActPlane~\cite{zheng2026actplane} / Tetragon~\cite{tetragonselectors} & & \checkmark & \checkmark & & \\
mcpguard-dynamic~\cite{mcpguarddynamic} & & \checkmark & \checkmark & & \\
ProcRoute~\cite{procroute2026} & & \checkmark & \checkmark & & \\
Claude Code / OpenShell~\cite{claudecodesandbox,openshell} & \checkmark & $\circ$ & & & \\
US\,7,080,408~\cite{US7080408B1} (quarantine) & \checkmark & & & \checkmark & \\
\textbf{Connection escrow} & \checkmark & \checkmark & \checkmark & \checkmark & \checkmark \\
\bottomrule
\end{tabular}
\end{table}

\section{Discussion}
\label{sec:discussion}
\paragraph{Where the cost actually is} The kernel hook does one map update and one address rewrite per connection; it is not the bottleneck at any scale we measured. The per-connection cost is a userspace round trip to the verifier and two signature operations, and the concurrency ceiling is the single-process Python interceptor. Both are implementation artifacts, not properties of the design: a compiled interceptor holding a persistent, pipelined connection to the verifier would cut the round trip to the verifier's own decision time (24\,$\mu$s measured) plus one network hop, and \texttt{SO\_REUSEPORT} workers on the shared maps scale the interceptor across cores. The design's intrinsic cost is one kernel map lookup plus whatever the policy itself takes.

\paragraph{The deadline is the real knob} For a workload whose connections are all policy-allowed, escrow adds a few milliseconds and is invisible against tool calls that take hundreds of milliseconds. The cost that bites is the \emph{hold} rule: a connection parked for a human or a slow external check pays up to the full deadline when no answer comes, and fails closed. Operators therefore reserve \texttt{hold} for genuinely ambiguous destinations and let the common case be a fast \texttt{allow}/\texttt{deny}; the calibrated-risk style of decision from our earlier gateway~\cite{zkproofgateway} fits here, deciding at once for the bulk and holding only the tail.

\paragraph{Composition} Escrow is not a network policy and not a mesh; it is a per-connection, identity-bound, fail-closed decision that sits underneath both. A cluster can run NetworkPolicy for coarse L3/L4 segmentation, a mesh for mTLS between services, and escrow for the agent-specific question of whether \emph{this} pod may open \emph{this} connection right now. Because it enforces below the application and below the sidecar, it also catches the traffic those layers miss: the child process, the raw socket, the static binary.

\paragraph{Toward a product} The natural next step is to pair the verdict-bound release demonstrated here with an attested record of the loaded program set, so a relying party can verify not only the decision but that the approved enforcement was in force when it was made --- the integrity half our earlier gateway~\cite{zkproofgateway} already provides. Escrow supplies the enforcement; that supplies the proof; together they are an auditable egress-governance layer for agent workloads.

\section{Limitations}
\label{sec:limits}
The hold point covers IPv4 TCP, and pins IPv4 UDP (including DNS) to a configured resolver through the \texttt{connect4} and \texttt{sendmsg4} hooks; IPv6 needs the \texttt{connect6} and \texttt{sendmsg6} hooks, which are not implemented. UDP is pinned rather than escrowed: a datagram to the cluster resolver is allowed and anything else denied, so DNS cannot be exfiltrated to an outside resolver, but a covert channel inside otherwise-allowed DNS traffic, and UDP payloads in general, are not held for a verdict. Relatedly, the agent resolves a name before it calls \texttt{connect()}, and a lookup of an attacker-controlled name reaches that name's authoritative server through the cluster resolver; escrow holds the connection, not the resolution, so G1 hides the connection attempt from the destination, not the fact that its name was looked up. The iptables fallback keys on uid and therefore assumes an unprivileged agent (Section~\ref{sec:holdpoint}); the eBPF path does not. A connection whose first bytes carry no SNI or recognisable request gives the verifier only the destination tuple. Attribution through \texttt{/proc} needs \texttt{CAP\_SYS\_PTRACE} on the node; the eBPF path avoids the scan for pid and cgroup but still reads \texttt{/proc/<pid>} for the executable digest. After the in-kernel rewrite the agent's \texttt{getpeername()} returns the interceptor's address; a \texttt{cgroup/getpeername4} program can present the original, and we have not written it. The prototype's interceptor is a single Python process and the verifier a threaded Python server; the concurrency figures reflect that, not the kernel program, and the paths to scale out are in Section~\ref{sec:discussion}. The evaluation is single-node; a multi-node run would exercise the DaemonSet and cross-node verifier traffic, which the manifests support but we have not measured. The content gate's uprobe covers shared and static OpenSSL and Go's \texttt{crypto/tls}; Rustls needs its own probe target, a stripped binary needs its offset supplied, the content model is a placeholder, each write is scored on up to one 16\,KiB record, and HTTP/2 headers need an HPACK decoder in the model; the gate is demonstrated on TLS~1.2 on loopback, not yet at cluster scale, and its adversary is hostile input to an honest agent, not hostile code (Section~\ref{sec:content}). The threat model trusts the node; an attacker with root on the node can detach the programs, which is why the product direction pairs escrow with attestation of the loaded set.

\section{Conclusion}
A connection can be held where it is born. Doing so in the kernel, attributing it to the socket and the pod, asking a verifier the agent cannot reach, and releasing exactly that socket once, gives an agent runtime a control that no library or proxy setting can route around, at a cost of a few milliseconds per connection. The decision logic that sits behind the verifier is a separate problem; this paper gives it a place to stand.

\section*{Reproducibility}
The implementation, the adversarial suite, the microbenchmark, the Kubernetes manifests and driver, and the logged runs are at \url{https://github.com/45h0kg/benign-attach-gateway}. \texttt{bash experiments/run\_all.sh} as root on a Linux 6.x host with cgroup v2, clang and iptables reproduces the suite and the microbenchmark; \texttt{experiments/k8s/drive.py}, \texttt{bypass.py} and \texttt{scale.py} reproduce the cluster results on a single-node Kubernetes host.

\end{document}

%% file: figures/tab_latency_body.tex
\begin{tabular}{lccc}
\toprule
Stage & eBPF, native & iptables, native & eBPF, pure Py \\
\midrule
Attribution (inet\_diag, /proc, registry) & 0.87 / 1.58 & 2.02 / 2.60 & 0.89 / 1.08 \\
Park first bytes & 0.10 / 0.15 & 0.11 / 0.14 & 0.10 / 0.14 \\
Verifier round trip (HTTP, policy, sign) & 2.21 / 3.51 & 2.44 / 3.02 & 3.69 / 4.90 \\
Release gate (verify, bind, consume) & 0.31 / 0.44 & 0.34 / 0.46 & 3.31 / 4.51 \\
Connect to destination & 0.43 / 0.60 & 0.47 / 0.62 & 0.53 / 0.74 \\
\textbf{Hold: accept to release} & 4.29 / 6.82 & 5.76 / 7.01 & 9.31 / 11.50 \\
Agent wall: connect, send, echo & 5.47 / 8.36 & 7.10 / 8.57 & 10.59 / 13.06 \\
Unsupervised baseline, agent wall & 0.74 / 0.96 & 0.86 / 1.20 & 0.84 / 1.12 \\
\midrule
16 agents: hold & 31.53 / 47.93 & 59.50 / 100.14 & 90.08 / 114.92 \\
16 agents: connections/s & 126.4 & 100.2 & 102.7 \\
\bottomrule
\end{tabular}